\documentclass[apj,numberedappendix,twocolappendix,appendixfloats]{emulateapj}

\tabletypesize{\scriptsize}  

\usepackage{rotating}
\usepackage{epsfig}

\usepackage[fleqn]{amsmath}

\newcommand{\lt}{\ifmmode\,<\,\else \,$<$\,\fi}
\newcommand{\kms}{\ifmmode\,{\rm km}\,{\rm s}^{-1}\else km$\,$s$^{-1}$\fi}

\newcommand{\magarc}{\ifmmode {{{{\rm mag}~{\rm arcsec}}^{-2}}}
             \else {{{mag}$~${arcsec}$^{-2}$}}
             \fi}

\newcommand{\funit}{\mathrm{ergs}~\mathrm{s}^{-1}~\mathrm{cm}^{-2}}
\newcommand{\Mstar}{\mathrm{M}_*}
\newcommand{\Msun}{\mathrm{M}_{\sun}}
\newcommand{\Lsun}{\mathrm{L}_{\sun}}
\newcommand{\Msunyr}{\mathrm{M}_{\sun}~\mathrm{yr}^{-1}}
\newcommand{\Hunit}{\mathrm{km}~\mathrm{s}^{-1}~\mathrm{Mpc}^{-1}}

\newcommand{\Hi}{H~{\sc i}}
\newcommand{\Hii}{H~{\sc ii}}

\newcommand{\Cii}{[C~{\sc ii}]}
\newcommand{\Oii}{[O~{\sc ii}]}
\newcommand{\Oiid}{[O~{\sc ii}]$\lambda\lambda$3727,3729}  
\newcommand{\Oiii}{[O~{\sc iii}]}
\newcommand{\Oiiit}{[O~{\sc iii}]$\lambda$4363}
\newcommand{\Oiiif}{[O~{\sc iii}]~52$\mu$m}
\newcommand{\Oiiie}{[O~{\sc iii}]~88$\mu$m}

\newcommand{\Pa}{Pa$\alpha$}
\newcommand{\Ha}{H$\alpha$}
\newcommand{\Hb}{H$\beta$}

\newcommand{\Sii}{[S~{\sc ii}]}

\newcommand{\Hp}{H$^{+}$}
\newcommand{\Op}{O$^{+}$}
\newcommand{\Opp}{O$^{++}$}
\newcommand{\Oppp}{O$^{3+}$}

\newcommand{\Te}{T$_{\mathrm e}$}
\newcommand{\Ne}{n$_{\mathrm e}$}
\newcommand{\Nunit}{cm$^{-3}$}
\newcommand{\OH}{$12 + \mathrm{\log{(O/H)}}$}
\newcommand{\OHpp}{$12 + \mathrm{\log{(O^{++}/H^{+})}}$}

\shorttitle{Direct Metallicities in the Reionization Epoch}

\begin{document}

\title{The Mass-Metallicity Relation at $z\simeq8$: \\
Direct-Method Metallicity Constraints and Near-Future Prospects}


\author{Tucker~Jones\altaffilmark{1}}
\author{Ryan~Sanders\altaffilmark{1,2}}
\author{Guido~Roberts-Borsani\altaffilmark{3}}
\author{Richard~S.~Ellis\altaffilmark{4}}
\author{Nicolas~Laporte\altaffilmark{5,6}}
\author{Tommaso~Treu\altaffilmark{3}}
\author{Yuichi~Harikane\altaffilmark{4,7}}

\affil{$^1$ 
Department of Physics, University of California Davis, 1 Shields Avenue, Davis, CA 95616, USA}
\affil{$^2$ 
Hubble Fellow}
\affil{$^3$ 
Department of Physics and Astronomy, University of California, Los Angeles, 430 Portola Plaza, Los Angeles, CA 90095, USA}
\affil{$^4$
Department of Physics and Astronomy, University College London, Gower Street, London WC1E 6BT, UK}
\affil{$^5$ 
Kavli Institute for Cosmology, University of Cambridge, Madingley Road, Cambridge CB3 0HA, UK}
\affil{$^6$ 
Cavendish Laboratory, University of Cambridge, 19 JJ Thomson Avenue, Cambridge CB3 0HE, UK}
\affil{$^7$
National Astronomical Observatory of Japan, 2-21-1 Osawa, Mitaka, Tokyo 181-8588, Japan}


\begin{abstract}

Physical properties of galaxies at $z>7$ are of interest for understanding both the early phases of star formation and the process of cosmic reionization. Chemical abundance measurements offer valuable information on the integrated star formation history, and hence ionizing photon production, as well as the rapid gas accretion expected at such high redshifts. We use reported measurements of \Oiiie\ emission and star formation rate to estimate gas-phase oxygen abundances in five galaxies at $z=7.1-9.1$ using the direct \Te\ method. We find typical abundances \OH~$=7.9$ ($\sim$0.2 times the solar value) and an evolution of $0.9\pm0.5$ dex in oxygen abundance at fixed stellar mass from $z\simeq8$ to 0. These results are compatible with theoretical predictions, albeit with large (conservative) uncertainties in both mass and metallicity. 
~
We assess both statistical and systematic uncertainties to identify promising means of improvement with the Atacama Large Millimeter Array (ALMA) and the James Webb Space Telescope (JWST). In particular we highlight \Oiiif\ as a valuable feature for robust metallicity measurements. Precision of 0.1--0.2 dex in \Te-based O/H abundance can be reasonably achieved for galaxies at $z\approx5$--8 by combining \Oiiif\ with rest-frame optical strong lines. It will also be possible to probe gas mixing and mergers via resolved \Te-based abundances on kpc scales. 
With ALMA and JWST, direct metallicity measurements will thus be remarkably accessible in the reionization epoch. 
\end{abstract}

\keywords{Galaxy chemical evolution (580) --- High-redshift galaxies (734) --- Galaxy evolution (594) --- Reionization (1383)}

\section{Introduction}\label{sec:intro}

Deep spectroscopy of distant galaxies has recently begun to provide both redshift confirmations and a glimpse of their physical properties at $z>7$. Multiple lines of evidence suggest that this period coincides with a phase transition of the universe from largely neutral at $z>7$ to primarily ionized at $z<6$, marking the ``epoch of reionization'' \citep[e.g.,][]{Stark2016,Mason2018}. A key question is whether star formation in early galaxies produced enough ionizing photons to account for the reionization process. 
Current constraints rely largely on the rest-frame ultraviolet luminosity function of photometrically-selected galaxy samples, measured from Hubble Space Telescope imaging. This provides a good instantaneous snapshot of star formation rates reaching $z\approx10$ \citep[e.g.,][]{Ellis2013,Calvi2016,Oesch2018,Bouwens2019}. However, information on the previous history of star formation and ionizing photon production remains ambiguous and challenging to obtain \citep[e.g.,][]{Strait2020,Roberts-Borsani2020}. Therefore relatively little is known about galaxy assembly in the first 500 Myr of the Universe at $z\gtrsim10$.

Gas-phase metallicity\footnote{Throughout this paper we use the term ``metallicity'' to refer to gas-phase abundance of oxygen relative to hydrogen, O/H.} 
of galaxies is a valuable diagnostic property as it is sensitive to star formation history, as well as cosmological gas accretion and metal-enriched outflows which are expected to be prevalent at high redshifts. A strong correlation between galaxy stellar mass and metallicity has been established up to $z\simeq3.5$, enabling interpretation of galaxy chemical evolution in terms of past star formation and gas flow rates \citep[e.g.,][and references therein]{Maiolino2019}. At higher redshifts, the widely used optical emission line diagnostics become inaccessible to ground-based observations, but will be observable with the upcoming James Webb Space Telescope (JWST). Currently only a few measurements of galaxy metallicities have been reported above $z>4$ using a variety of methods \citep[e.g.,][]{Shapley2017,Totani2006,Faisst2016,Cullen2019}.

This paper is concerned with direct measurement of galaxy metallicities in the epoch of reionization. 
The standard ``direct method'' (or ``\Te\ method'') to calculate metallicity relies on nebular electron temperature \Te\ and density \Ne. \Te\ is the more challenging to obtain and is typically determined from ratios of auroral and strong nebular emission lines (e.g., \Oiiit/\Oiii$\lambda$5007). However, auroral line fluxes are well below the detection thresholds of typical high-redshift surveys. To date this method has been applied for tens of galaxies at $z\sim1$ \citep[e.g.,][]{Jones2015,Ly2014} and only a handful at $z>2$ \citep[e.g.,][]{Sanders2020, Christensen2012, Gburek2019}, with none at $z>4$. Applying the \Te\ method with optical auroral lines at higher redshifts will be extremely challenging even with JWST. In contrast, several detections of far-IR \Oiiie\ emission have recently been achieved at $z>7$ using the Atacama Large Millimeter Array \citep[ALMA;][]{Inoue2016,Laporte2017,Carniani2017,Hashimoto2018,Hashimoto2019,Tamura2019}. 
This breakthrough with ALMA provides a promising route toward direct metallicity measurements of bright $z>7$ galaxies using \Oiiif,88$\mu$m combined with \Oiii$\lambda\lambda$4959,5007 to determine \Te. In fact the \Te\ derived using far-IR lines is less sensitive to variations in temperature than auroral diagnostics, potentially enabling more robust results \citep[e.g.,][]{Croxall2013,Esteban2009}.

This paper addresses both available constraints and future prospects for galaxy metallicities at $z>7$. We first describe our method and apply it to a sample of $z>7$ galaxies (Section~\ref{sec:sample}), followed by an assessment of systematic uncertainties (Section~\ref{sec:uncertainties}). We discuss implications of our results for chemical evolution in the reionization epoch in Section~\ref{sec:discussion}. In Section~\ref{sec:jwst} we consider how to mitigate uncertainties with future observations using JWST and ALMA, feasibly reaching 0.1--0.2 dex precision in O/H. We summarize the main conclusions in Section~\ref{sec:summary}. 
Throughout this paper we adopt a flat $\Lambda$CDM cosmology with $\Omega_M=0.272$, $\Omega_{\Lambda}=0.728$, and H$_0=70.4~\Hunit$ \citep{Komatsu2011}. 
We refer to emission lines by their rest-frame wavelengths, adopting common but separate conventions for optical (e.g., \Oiiit, implicitly in Angstroms) and infrared lines (e.g., \Oiiie).

\section{Metallicity constraints at $z>7$ from current \Oiiie\ data}\label{sec:sample}

In this section we show that a combination of \Oiiie\ nebular emission and star formation rate (SFR) alone provides reasonable constraints on the metallicity, subject to assumptions about nebular physical properties. The basic methodology is as follows:
\begin{enumerate}
\item Estimate \Hi\ recombination line luminosity (e.g. \Hb) based on the photometrically-derived SFR
\item Calculate the abundance ratio \Opp/\Hp\ from the ratio of \Oiiie/\Hb\ luminosities, with an assumed electron density (\Ne) and temperature (\Te)
\item Assess systematic uncertainties in abundance based on a plausible range of \Ne, \Te, and ionization correction factor (ICF)
\end{enumerate}
A notable aspect is that \Te\ contributes relatively little uncertainty, whereas it typically dominates the error budget of \Oiiit-based measurements. This advantage results from the relative insensitivity of far-IR emission lines to temperature (as we discuss in Section~\ref{sec:52um}).

\subsection{$z>7$ galaxy sample}

We analyze the sample of six Lyman break galaxies with \Oiiie\ detections compiled by \citet{Harikane2020}, spanning $z=7.1$--9.1. The relevant measurements and original literature references are given in Table~\ref{tab:metallicity}. In the case of BDF-3299, the SFR and \Oiiie\ measurements correspond to different spatial regions which we consider separately, with appropriate upper and lower limits. The sample therefore contains 7 sources, of which 5 have measurements of both SFR and \Oiiie. 

We estimate \Hi\ recombination line luminosities using the calibration of \citet{Kennicutt1998} converted to a Chabrier stellar initial mass function (IMF):
\begin{equation}
L(\mathrm{H\alpha})~\mathrm{[erg~s^{-1}]} = 2.3\times10^{41} ~ \times ~ \mathrm{SFR~[\Msunyr]} 
\label{eq:L_Ha}
\end{equation}
as appropriate for the SFR values in Table~\ref{tab:metallicity} \citep[for details see][and other tabulated references]{Harikane2020}. This calibration is consistent (within 5\%) with the mean relation for $z\sim2$ galaxies from \citet{Shivaei2016}. We report \Hb\ luminosity as a practical example (e.g., easily compared with \Oiii$\lambda\lambda$4959,5007), with intrinsic ratio $\mathrm{\frac{H\alpha}{H\beta}} = 2.79$ appropriate for Case B recombination and \Te~$=1.5\times10^4$~K. Both the SFR and luminosity values in Table~\ref{tab:metallicity} are corrected for dust attenuation (with a range $A_V \approx 0$--1 from original references), while observed \Hb\ fluxes are expected to be reduced.

\subsection{Oxygen abundance}\label{sec:abundance_estimates}

The combination of \Oiiie\ and \Hb\ luminosities now allows an estimate of the ion abundance ratio \Opp/\Hp, which we compute using the {\sc PyNeb} package \citep{Luridiana2012}. 
We adopt fiducial values of \Te~$= 1.5\times10^4$~K and \Ne~$= 250$~\Nunit, which gives a simple relation for doubly ionized oxygen abundance:
\begin{equation}
12+\log{(\mathrm{O^{++}/H^+})} = 7.735 + \log{\frac{\mathrm{L_{88 \mu m}}}{\mathrm{10^7~\Lsun}}} - \log{\frac{\mathrm{SFR}}{\mathrm{\Msunyr}}}
\label{eq:Opp_abundance}
\end{equation}
expressed here in terms of SFR. 
These fiducial values are chosen on the basis of available measurements for samples at $z\gtrsim2$ \citep[e.g.,][]{Sanders2016,Sanders2020,Strom2017}; both the fiducial \Te\ and \Ne\ are considerably higher than for typical galaxies at $z\simeq0$ \citep[e.g.,][]{Andrews2013}. 
The $z\gtrsim2$ sample with temperature measurements available is in fact representative of even higher-$z$ galaxy demographics in terms of specific SFR, young inferred ages, emission line equivalent widths, and other properties. 
Resulting ion abundance ratios are given in Table~\ref{tab:metallicity}, along with total abundance \OH\ for an estimated ionization correction of 0.17 dex (Section~\ref{sec:ICF}). 

This simple exercise demonstrates the method of determining oxygen abundances, with fiducial assumptions suggesting approximately 0.05--0.4 times the solar value for this $z>7$ sample \citep[adopting solar \OH~$=8.69$;][]{Asplund2009}. However we must first verify the method (Section~\ref{sec:DGS}) and critically assess the uncertainties (Section~\ref{sec:uncertainties}), before discussing implications of the results.

\subsection{Validation of the method at $z=0$}\label{sec:DGS}

\begin{figure}
\centerline{
\includegraphics[width=\columnwidth]{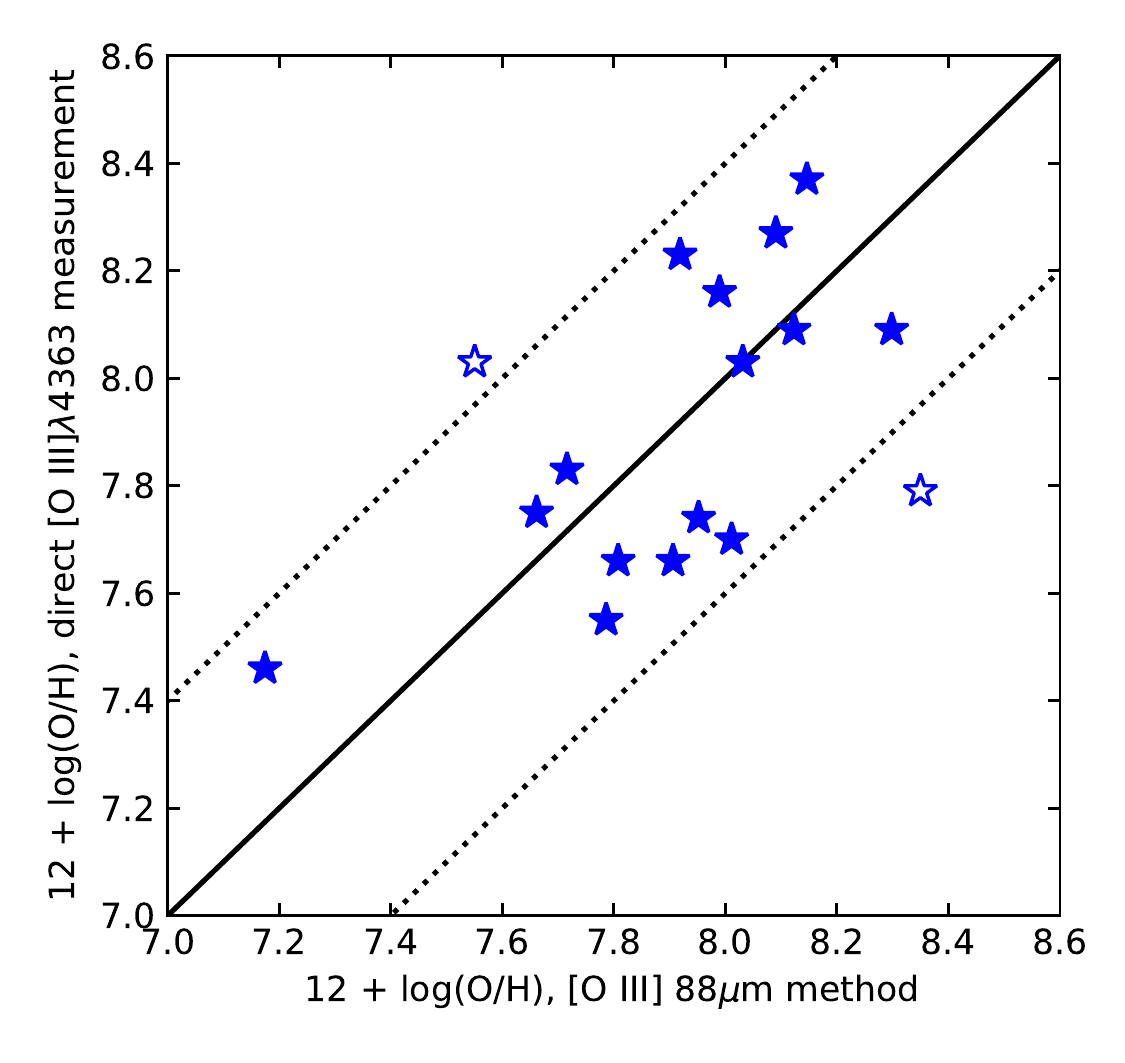}
}
\caption{
\label{fig:DGS}
Comparison of \OH\ determined from \Oiiie\ and SFR, compared to measurements based on standard optical emission line methods (using \Oiiit\ as a \Te\ diagnostic). Blue stars show all galaxies with suitable \Oiiie\ measurements from Herschel taken as part of the Dwarf Galaxy Survey (DGS). 
The DGS sample probes sub-solar metallicities relevant for comparison to high redshift galaxies (here we adopt a solar abundance \OH~$=8.69$). 
The solid black line shows the one-to-one relation, with dashed lines at $\pm$0.4 dex (corresponding to 2$\sigma$ scatter of the solid symbols), demonstrating reasonably good agreement. 
We note that the dashed lines serve to illustrate scatter about the one-to-one relation and are not intended to represent confidence intervals.
The galaxies which differ by $>$0.4 dex (open symbols) can be reconciled with direct measurements as discussed in the text. 
}
\end{figure}

To verify the reliability of using \Oiiie\ and SFR to calculate metallicity, we apply the method described above to nearby galaxies with Herschel Space Observatory data from the Dwarf Galaxy Survey \citep[DGS;][]{Madden2013}. The DGS sample is chosen as being likely the most appropriate comparison based on properties such as nebular excitation and metallicity. 
Table~\ref{tab:DGS} lists relevant physical properties for all 19 DGS targets for which infrared-based star formation rates are available \citep{Madden2013}, and for which Herschel spectroscopy covers the entire far-IR emission region. 
Figure~\ref{fig:DGS} shows the resulting metallicities calculated using \Oiiie\ fluxes from \cite{Cormier2015}, with distances and SFRs from \cite{Madden2013}, as given in Table~\ref{tab:DGS}. These \Oiiie-based metallicities are compared with optical \Oiiit-based direct measurements, as reported by \citet{Madden2013} for 17 of the 19 galaxies (using the method of \citealt{Izotov2006}; Table~\ref{tab:DGS}). 
For comparison purposes we subtract 0.2 dex from the \Oiiie-based metallicities in Figure~\ref{fig:DGS}, to account for systematic differences in the \Oiiit\ method attributed to temperature fluctuations (i.e., the well-known ``abundance discrepancy factor'' or ADF; \citealt{Esteban2009, Lopez-Sanchez2012, Blanc2015}). \Oiiie\ is not significantly affected by such temperature fluctuations, although it may be subject to other biases. We have otherwise adopted the same fiducial values as for the $z>7$ sample described above.

Figure~\ref{fig:DGS} shows generally good agreement between our method and previously published O/H values (Table~\ref{tab:DGS}), with RMS scatter of 0.26 dex. The average offset is $<$0.01 dex, although a systematic offset can easily be introduced (or removed) by varying the fiducial assumptions such as density, ionization correction, or the ADF. 
It is notable that the scatter in Figure~\ref{fig:DGS} is $<$0.3 dex given that a single \Te, \Ne, and ICF has been assumed for the entire DGS sample. Only two of the 17 galaxies in Figure~\ref{fig:DGS} are discrepant by $>$0.35 dex, and we consider these cases specifically:
\begin{itemize}
\item {\it Mrk 1450}: Metallicity from the fiducial \Oiiie\ method is lower by $\Delta(\log{\textrm{O/H}})=-0.48$ than the \Oiiit-based value reported in \cite{Madden2013}. This discrepancy appears to arise largely from the SFR, which is used only as a proxy for \Hi\ line luminosity. The extinction-corrected \Hi\ Balmer line flux \citep{Izotov1994} corresponds to SFR~$=0.09\, \Msunyr$ (via Equation~\ref{eq:L_Ha}), comparable to the SFR~$=0.11 \, \Msunyr$ reported by \citet{Sargsyan2009} based on 1.4 GHz measurements, and lower than the 0.43~$\Msunyr$ from \citet{Madden2013}. 
Adopting this direct measurement of \Hi\ line luminosity along with \Te, \Ne, and ICF \citep{Izotov2014,Izotov1994}, we find good agreement with $\Delta(\log{\textrm{O/H}})=0.02$. 

\item {\it UM 461}: The difference $\Delta(\log{\textrm{O/H}})=0.56$ cannot be explained by \Ne\ and \Te, which are both similar to the fiducial values \citep{Lagos2018}. However, \cite{Lagos2018} report an \Ha-based SFR~$=0.077~\Msunyr$ (indicating higher \Hi\ luminosity c.f. $0.01~\Msunyr$ from \citealt{Madden2013}) and a relatively low ICF of 0.03 dex. Adopting these measurements, the difference is $\Delta(\log{\textrm{O/H}})=-0.14$. 
\end{itemize}
In sum, the largest outliers from Figure~\ref{fig:DGS} are in fact consistent ($<0.2$ dex) with the \Oiiie\ method when direct measurements are used instead of fiducial values. The RMS scatter in Figure~\ref{fig:DGS} is only 0.2 dex if these two galaxies are removed or corrected; these cases serve to illustrate potential causes and rates of outliers. The data show a positive correlation at 97\% confidence level (Pearson correlation coefficient $\rho=0.53$), or 99.7\% confidence ($\rho=0.70$) if these two galaxies are excluded. 

We have also performed the comparison shown in Figure~\ref{fig:DGS} with metallicities derived from the \citet{Pilyugin2005} method \citep[Table~\ref{tab:DGS}; for a detailed comparison see Appendix~A of][]{Madden2013}. This gives nearly identical results with an RMS scatter of 0.30 dex (0.20 dex with outliers removed), and positive correlation at 95\% confidence level (99.9\% with outliers removed), although the metallicities from this method are on average 0.09 dex higher. The three cases which differ by $>$0.35 dex are all reconciled (within $<$0.2 dex) when direct measurements of \Te, \Ne, and ICF are used instead of fiducial values. The main conclusions of this comparison are therefore unchanged regardless of which metallicity values in Table~\ref{tab:DGS} \citep[via][]{Madden2013} are adopted.

These results demonstrate that the combination of \Oiiie\ and SFR alone can empirically reproduce standard \Te-method metallicity measurements to within $\sim$0.3 dex for individual galaxies in the DGS sample, and $<0.2$ dex using prior information (such as ICF and \Ne). We conclude that the basic method presented here is sound.

\subsection{The advantage of \Oiiif, 88$\mu$m over \Oiiit}\label{sec:52um}

\begin{figure}
\centerline{
\includegraphics[width=\columnwidth]{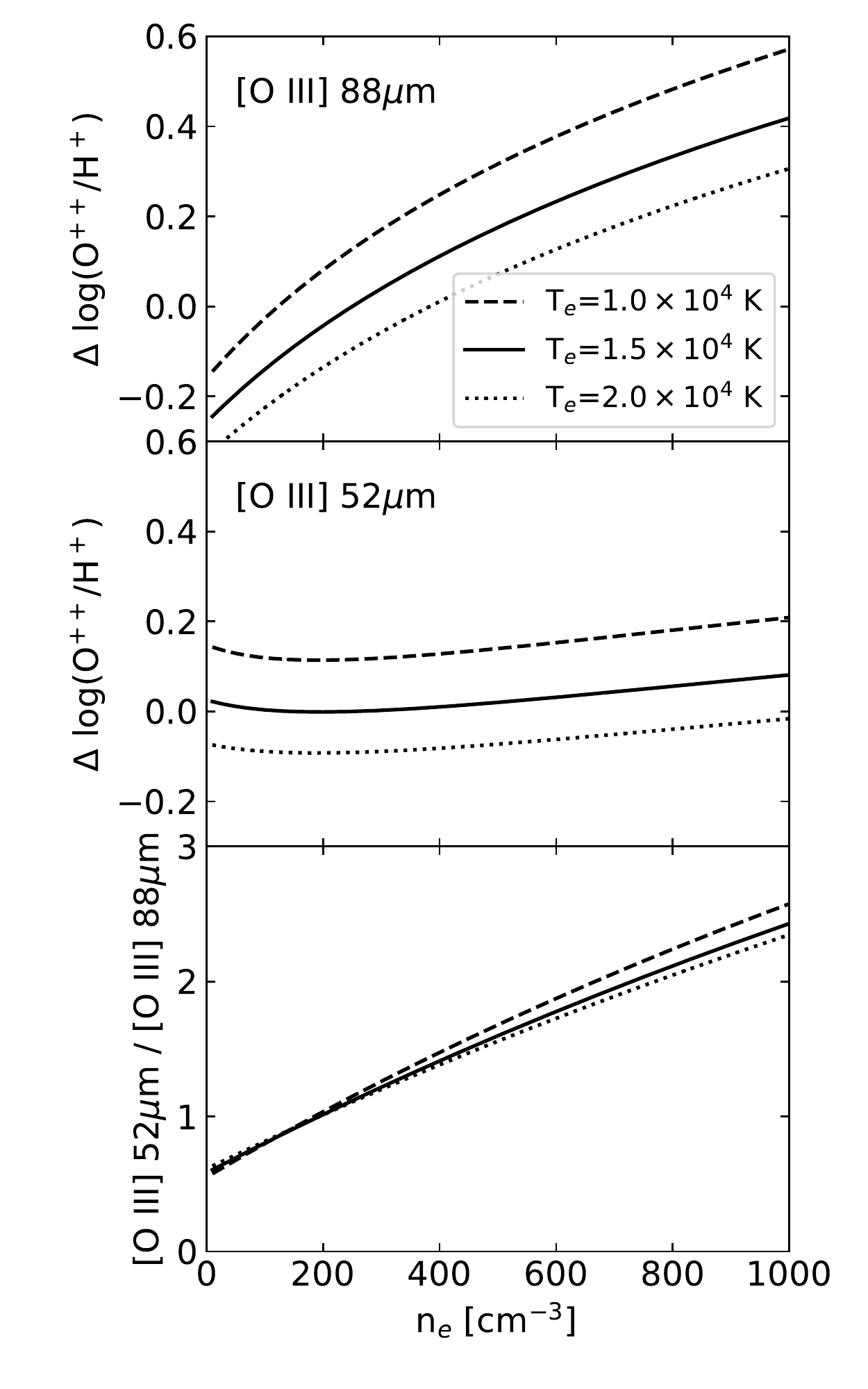}
}
\caption{
\label{fig:density}
Illustration of how assumed density and temperature values affect the derived metallicity. 
The top panel shows how derived \Opp\ abundance varies relative to the fiducial values (Section~\ref{sec:sample}) at fixed \Oiiie\ flux. A strong dependence on \Ne\ is apparent due to suppression of \Oiiie\ emission by collisional de-excitation. 
In contrast, the middle panel shows that \Oiiif-based abundance is nearly independent of density for \Ne~$\lesssim1000$~\Nunit, with only modest temperature dependence. \Oiiif\ is therefore a promising metallicity diagnostic for future work. 
The bottom panel demonstrates how the \Oiii\ doublet ratio is a robust density diagnostic \citep[e.g.,][]{Palay2012}, with \Oiiif\ becoming the stronger line at \Ne~$>200$~\Nunit. 
}
\end{figure}

A principal motivation for the \Oiiie-based approach is the demonstrated success in detecting far-IR lines at $z>7$. It is not a very pragmatic method at low redshift, where \Oiiit\ is instead the most common \Te\ diagnostic. This is in part due to challenges of infrared observations and the relative ease of detecting optical auroral lines in nearby galaxies. However, the situation is reversed at high redshifts where far-IR lines become accessible to ALMA. A key advantage of far-IR lines is their relative brightness compared to auroral lines. Another major advantage noted previously is that far-IR lines are relatively insensitive to temperature \citep[e.g.,][]{Croxall2013}. Over a range of 9,000--20,000~K, \Oiiie\ varies by only 15\% in emissivity, whereas \Oiiit\ varies by a factor 33$\times$. This causes \Oiiit\ and other auroral line measurements to be strongly biased toward high-temperature regions. As a result, current $z\gtrsim1$ auroral line samples are a biased subset of the overall galaxy population \citep[e.g.,][]{Sanders2020}. In contrast, far-IR lines depend primarily on ion abundances and thus can provide more robust metallicities across a broader sample. 

While this work is necessarily limited to available measurements of the 88~$\mu$m line, we note that \Oiiif\ is a superior diagnostic. 
Figure~\ref{fig:density} demonstrates that both lines have similarly weak temperature dependence, but \Oiiif\ is far less sensitive to collisional de-excitation at densities \Ne~$\lesssim1000$~\Nunit, and thus enables considerably more robust metallicity measurements. Moreover the \Oiiif\ line is comparable or brighter (with minimum flux ratio $\frac{f_{52}}{f_{88}}\geq 0.6$; Figure~\ref{fig:density}). 
Therefore we consider observations of the 52~$\mu$m line to be a highly desirable priority for future work. Unfortunately it is not widely available even for nearby galaxies, and hence our team is pursuing observations for a benchmark sample with SOFIA (the Stratospheric Observatory For Infrared Astronomy).

\section{Sources of uncertainty}\label{sec:uncertainties}

The scatter in Figure~\ref{fig:DGS} and case studies discussed in Section~\ref{sec:DGS} underscore potential systematic errors arising from the limited data currently available for $z>7$ galaxies. This section focuses on assessing the main sources of systematic uncertainty: nebular temperature \Te, density \Ne, ionization correction factor (ICF), and \Hi\ line luminosity. Each of these contributes at the level of $\sim$0.2 dex. 
Our approach is to consider the plausible range of values for each parameter in the $z>7$ sample, which we list in Table~\ref{tab:uncertainty} along with the corresponding range of metallicity. We also tabulate the partial derivative of O/H with respect to each parameter, illustrating the sensitivity to moderate changes in the assumed values. 
This reveals the dominant sources of error and consequently {\it how to minimize uncertainty} for future studies. The effects of \Te\ and \Ne\ are illustrated as an example in Figure~\ref{fig:density}. 
Below we discuss each parameter separately.

\subsection{Nebular temperature \Te}

Higher \Te\ results in lower metallicity, although with modest dependence. We consider a maximal range of \Te(\Opp)~=~9,000--20,000 K. This spans the full range of nebular \Te\ values seen in galaxy surveys \citep[e.g.,][]{Izotov2006,Andrews2013}, hence we view this as highly conservative. Allowing values as low as 5,000 K would increase the range by only 0.07 dex above the nominal bound. Restricting to 11,000--20,000 K, the range in O/H would decrease to $\pm0.1$ dex. 
Thus the \Opp\ abundance can be reasonably constrained even with completely unknown \Te.

\subsection{Nebular density \Ne}

Higher \Ne\ results in higher metallicity, due to collisional de-excitation suppressing the \Oiiie\ emission (Figure~\ref{fig:density}). Our lower bound of 10~\Nunit\ represents a strict limit; lower densities would change O/H by only $\sim$1\%. The upper bound of 600~\Nunit\ is the approximate maximum range found at $z\sim2.3$ from the MOSDEF survey \citep[from \Oii\ and \Sii;][]{Sanders2016}. Higher densities are inferred for some objects but not at a statistically significant level. Nonetheless, higher densities are possible (e.g., \Ne~$>1000$~\Nunit\ in extreme $z\sim3$ starbursts; \citealt{Zhang2018}) in which case true metallicities would be above the nominal bounds.

\subsection{Ionization correction factor}\label{sec:ICF}

Only the \Opp\ ion is directly constrained with present data, representing a lower limit on total oxygen abundance. To estimate the ICF we first adopt a fiducial reddening-corrected $\mathrm{O_{32}} = \frac{\text{\Oiii} \lambda\lambda 4959,5007}{\text{\Oiid}}$ flux ratio which is readily compared with observations. 
High \Oiiie\ fluxes and \Oiii/\Cii\ ratios generally suggest high $\mathrm{O_{32}}$ for the $z>7$ sample \citep[e.g.,][]{Harikane2020}. This is supported by observed mid-IR colors of $z>6$ galaxies indicating extremely large \Oiii\ equivalent widths \citep[typically W$_{\mathrm{[O III] 4959,5007 + H\beta}} \approx 500-1500$ \AA; e.g.,][]{Labbe2013, Smit2014, Smit2015, Laporte2014, Huang2016, Roberts-Borsani2016}, confirmed in lower-redshift analogs \citep[e.g.,][]{Mainali2020}. 
Empirically such high equivalent widths suggest $\mathrm{O_{32}} \simeq 2$--10 \citep[accounting for scatter in the relation;][]{Sanders2020}. However this could be overestimated if there are significant Balmer breaks in the stellar spectra. 
Theoretical modeling likewise suggests large ionization parameters and high $\mathrm{O_{32}}$. For example, \cite{Katz2019} predict $\mathrm{O_{32}} \simeq 3$--8 for simulated massive galaxies at $z\simeq10$. 

Motivated by these expectations we adopt a fiducial $\mathrm{O_{32}}=3$, combined with \Ne~$=250$~\Nunit\ and \Te(\Op) based on an \Hii\ region relation: 
$$ \textrm{\Te(\Op)} = 0.7 \times \textrm{\Te(\Opp) + 3000 K} $$ 
(the ``$T_2-T_3$ relation''; \citealt{Campbell1986,Stasinska1982}. An empirical alternative \Te(\Op)~=~\Te(\Opp)$-1300$ K from \citealt{Andrews2013} produces very similar results.). 
This gives 33\% of oxygen in the \Op\ state, or an ICF of $+0.17$ dex from \Opp\ to total O abundance.
As an approximate upper bound we take $\mathrm{O_{32}}=1$, \Ne~$=600$ \Nunit, and \Te(\Opp)~=~9,000 K, giving ICF~$=0.39$ dex. As a lower bound we take $\mathrm{O_{32}}=10$, \Ne~$=10$ \Nunit, and \Te(\Opp)~=~\Te(\Op)~=~20,000 K, giving ICF~$=0.04$ dex. 
These cases span the maximal ranges considered for \Te\ and \Ne. However, lower $\mathrm{O_{32}}$ or \Te(\Op) values are possible and would lead to higher O/H (e.g., for He 2-10 discussed in Section~\ref{sec:DGS}). 

The above estimates consider only singly and doubly ionized oxygen. Neutral ions can be safely ignored for \Hii\ regions. The \Oppp\ state may contribute in cases where \Opp/\Op\ is very large, but the expected effect is of order $\sim$1\% for the conditions described above \citep[e.g.,][]{Izotov2006,Guseva2012}. If anything, \Oppp\ would contribute in more extreme cases of small ICF where \Op\ has been underestimated, offsetting the overall effect. Hence we consider the ICF from \Oppp\ and higher ionization states to be negligible.

\subsection{Hydrogen recombination line luminosity}

Higher \Hi\ recombination line luminosity implies higher \Hp\ abundance and lower O/H. We estimate uncertainty using the comparison of \Hi\ Balmer emission with photometrically-derived SFR by \cite{Shivaei2016}, who find a scatter of $\sigma=0.17$--0.26 dex for various data sets in their $z\sim2$ sample. Therefore we adopt a somewhat conservative 0.25 dex systematic uncertainty in \Hi\ line luminosity (e.g., \Hb). This is comparable in many cases to statistical uncertainty in SFR of the $z>7$ sample, whose spectral energy distributions (SEDs) are less well sampled than galaxies at lower redshift.

\subsection{Total systematic uncertainty}

We now assess the total systematic uncertainty, noting that the various contributions are not fully independent. Higher \Te\ decreases both \Opp/\Hp\ and the ICF toward lower O/H. Higher \Ne\ increases \Opp/\Hp\ and slightly increases the ICF. \Te\ also correlates with \Ne\ (i.e., higher overall pressure), such that their combined effects may lead to somewhat lower systematic uncertainty in O/H. Overall we estimate the combined uncertainty from \Te, \Ne, and ICF as $\simeq0.3$ dex considering their covariances. 
Scatter in the SFR--$L$(\Hb) relation has no clear connection to the other parameters and we treat it as independent. In combination we arrive at an estimate of $\sigma_{sys}(\log{\mathrm{(O/H)}})=0.4$ dex for the {\it total systematic error} in O/H for the $z>7$ sample. We reiterate that this estimate conservatively spans a broad range of physical parameters (e.g., \Te~=9,000--20,000 K). 
For comparison, current measurements based on \Oiiit\ at $z\gtrsim2$ have typical statistical uncertainties of $\sim$0.2 dex in O/H \citep[e.g.,][]{Sanders2020}, comparable to the scatter from applying our method to the DGS sample (Figure~\ref{fig:DGS}). In our view it is remarkable that such precision can be achieved at $z>7$ solely from \Oiiie\ flux and a photometry-based estimate of the SFR.

\section{Chemical enrichment at $z\simeq8$}\label{sec:discussion}


Having verified our methodology and assessed the uncertainties, we now discuss our results for metallicity of the $z>7$ sample. First we consider timescales required for enrichment and implications for extended star formation histories, followed by a discussion of the mass-metallicity relation and its evolution to the present day. We then comment on spatially resolved structure in the sample.

\subsection{Enrichment and star formation timescales}

The metallicity estimates in Table~\ref{tab:metallicity} range from $\sim\frac{1}{20}$ to $\frac{1}{3}$ the solar value, albeit with uncertainty spanning $\sim$1 dex. The average is \OH~$=7.94$ or $\sim$0.2 solar. 
Notably in two cases the 1-$\sigma$ lower bounds are $\gtrsim0.1$ solar, indicating a significant degree of chemical enrichment only $\sim$650 Myr after the big bang. 
We can place an approximate lower bound on the enrichment timescale by considering a closed-box chemical evolution model, with an estimated gas depletion time $\tau\approx300$~Myr at these redshifts \citep[e.g.,][]{Scoville2017}. 
For a yield of approximately solar abundance \citep[e.g.,][]{Woosley1995}, a metallicity of 0.1 solar requires that $\sim$10\% of gas in these systems has already been processed into stars, requiring $\sim$30 Myr of constant star formation. The enrichment timescale could be reduced if oxygen yields are higher, as may be expected at low metallicity due to stellar evolution effects and possibly a top-heavy IMF (e.g., \citealt{Jerabkova2018}). With yields from low-metallicity stars $y_O=0.007$--0.039, as tabulated by \citet[][for a Chabrier IMF of different mass cutoffs]{Vincenzo2016}, an abundance of 0.1 solar is reached in only 6--33 Myr for a closed-box model. 
However, gas inflows and outflows would increase the time required to enrich interstellar gas. Both effects are expected to be prevalent given the sample's high SFRs. As an example, \cite{Langan2020} report that gas flows reduce effective yields by a factor of 10 in simulated $z\approx8$ galaxies. This drives the timescale to $\sim$100--300~Myr to reach enrichment of $\sim$0.1 solar. 

We thus view 100 Myr as a reasonable order-of-magnitude estimate of the timescale of past star formation in the $z>7$ sample, in order to explain the average metallicities. 
Ages of $\sim$10 Myr are plausible only in the absence of gas flows, which we view as an unlikely scenario, or for the lower-metallicity galaxies. 
This contrasts with ages $\lesssim10$ Myr inferred for most of the sample from single stellar population SED fits. Our chemical evolution results thus indicate that the true stellar mass is likely dominated by an extended period of star formation which is not captured by single-component photometric models \citep[e.g.,][]{Hashimoto2018,Roberts-Borsani2020}, especially given limited available data beyond the rest-frame ultraviolet (UV). 
These results are generally consistent with previous metallicity estimates, for example of SXDF-NB1006-2 \citep[0.05--1 solar;][]{Inoue2016} and MACS0416-Y1 \citep[$\sim$0.2 solar;][]{Tamura2019}, where in both cases the authors conclude that dust and metal enrichment requires the presence of an evolved underlying stellar population.

A key diagnostic in determining the extent of previous star formation is the strength of the stellar Balmer break. However, this feature is challenging to distinguish from strong nebular emission in photometry of the $z>7$ sample. Red mid-IR 3.6$\mu$m - 4.5$\mu$m ``IRAC excess'' colors measured with the Spitzer Space Telescope can be explained by very young stellar populations with strong nebular emission lines, or by more extended star formation histories with weaker emission lines and a prominent Balmer break \citep[permitting $\sim$30 times higher stellar mass in the latter case;][]{Roberts-Borsani2020}. 

The strongest indications of extended star formation based on chemical evolution are for MACS1149-JD1 and B14-65666, where our method indicates relatively high metallicity. The lower bound on metallicity is $\gtrsim$0.1 solar in both cases, considering both systematic and measurement uncertainties. Such enrichment suggests previous star formation lasting several tens of Myr or more as discussed above. This result bolsters previous interpretation of MACS1149-JD1 where the IRAC excess cannot be explained by line emission, clearly favoring a Balmer break \citep[with stellar population age $\gtrsim100$ Myr;][]{Hashimoto2018}. 

In the case of B14-65666, the photometric data are consistent with a young starburst of age $\lesssim$10 Myr and strong nebular emission \citep{Hashimoto2019}, but the data also permit extended star formation with a moderate Balmer break \citep{Roberts-Borsani2020}. 
Such a young single-starburst age would imply extremely high effective yields to reach the metallicity we derive. A more likely scenario in our view is that the gas may have been pre-enriched by previous generations of star formation, which do not necessarily dominate the observed SED nor the IRAC excess, but can nonetheless dominate the total stellar mass and chemical enrichment. 
This is demonstrated by the 2-component SED fit of \cite{Roberts-Borsani2020}: a young (3 Myr) starburst contributes most of the broadband photometric flux, while an extended star formation component constitutes 92\% of the stellar mass (and we note that this model provides the best fit to the data based solely on log-likelihood). 

Thus our metallicity estimates provide additional constraints with which to determine the star formation histories of high-z galaxies, despite some ambiguity in photometric analyses. 
This supports the general results of \cite{Roberts-Borsani2020} that current near- and mid-IR data permit a broad range of star formation histories and stellar masses, which are not captured by single-component models. 
While JWST spectroscopy will directly resolve this issue, we conclude that chemical enrichment constraints from ALMA are valuable in support of interpreting stellar population synthesis models.

\subsection{The $z\simeq8$ mass-metallicity relation}\label{sec:MZR}

\begin{figure*}
\centerline{
\includegraphics[width=\columnwidth]{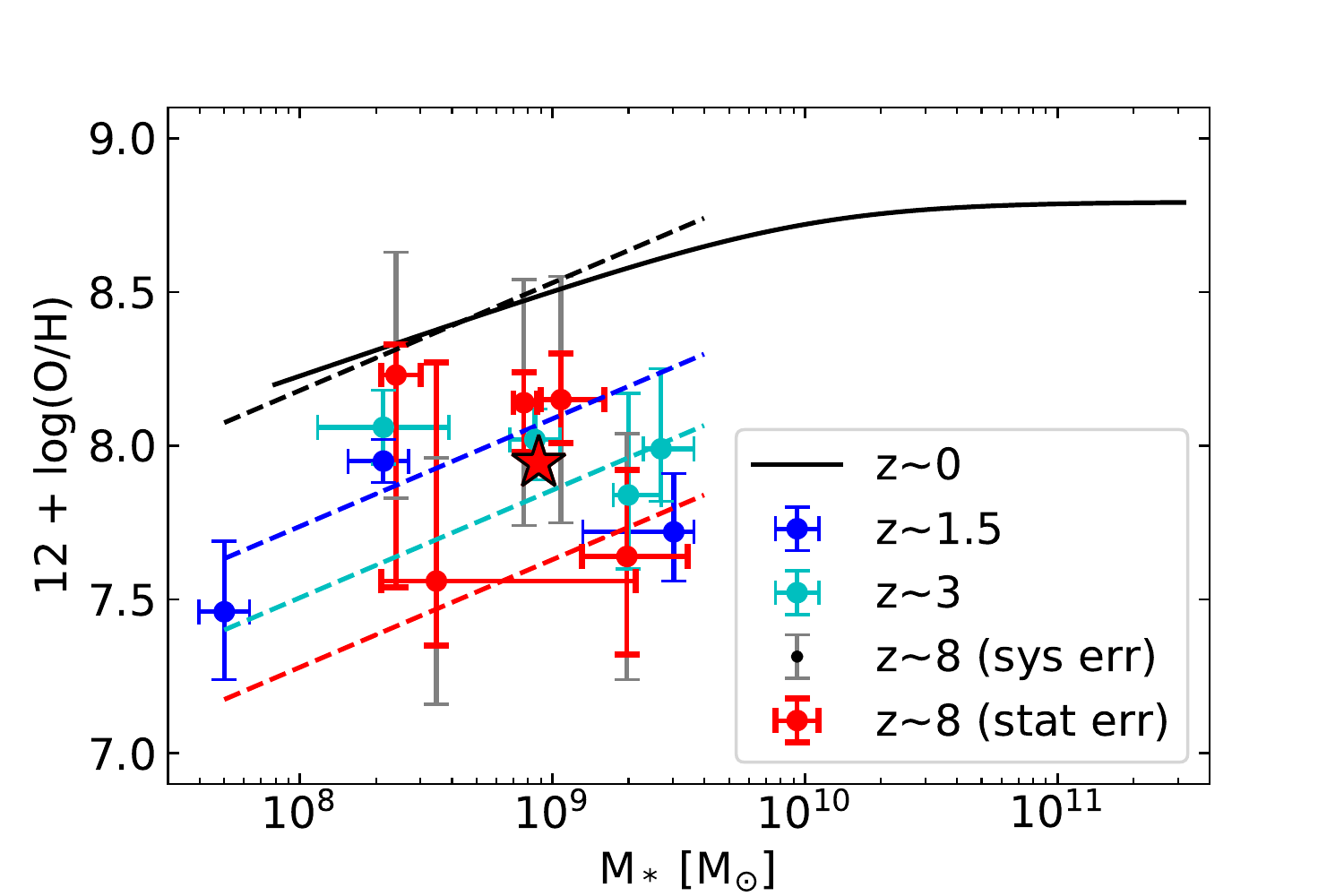}
\includegraphics[width=\columnwidth]{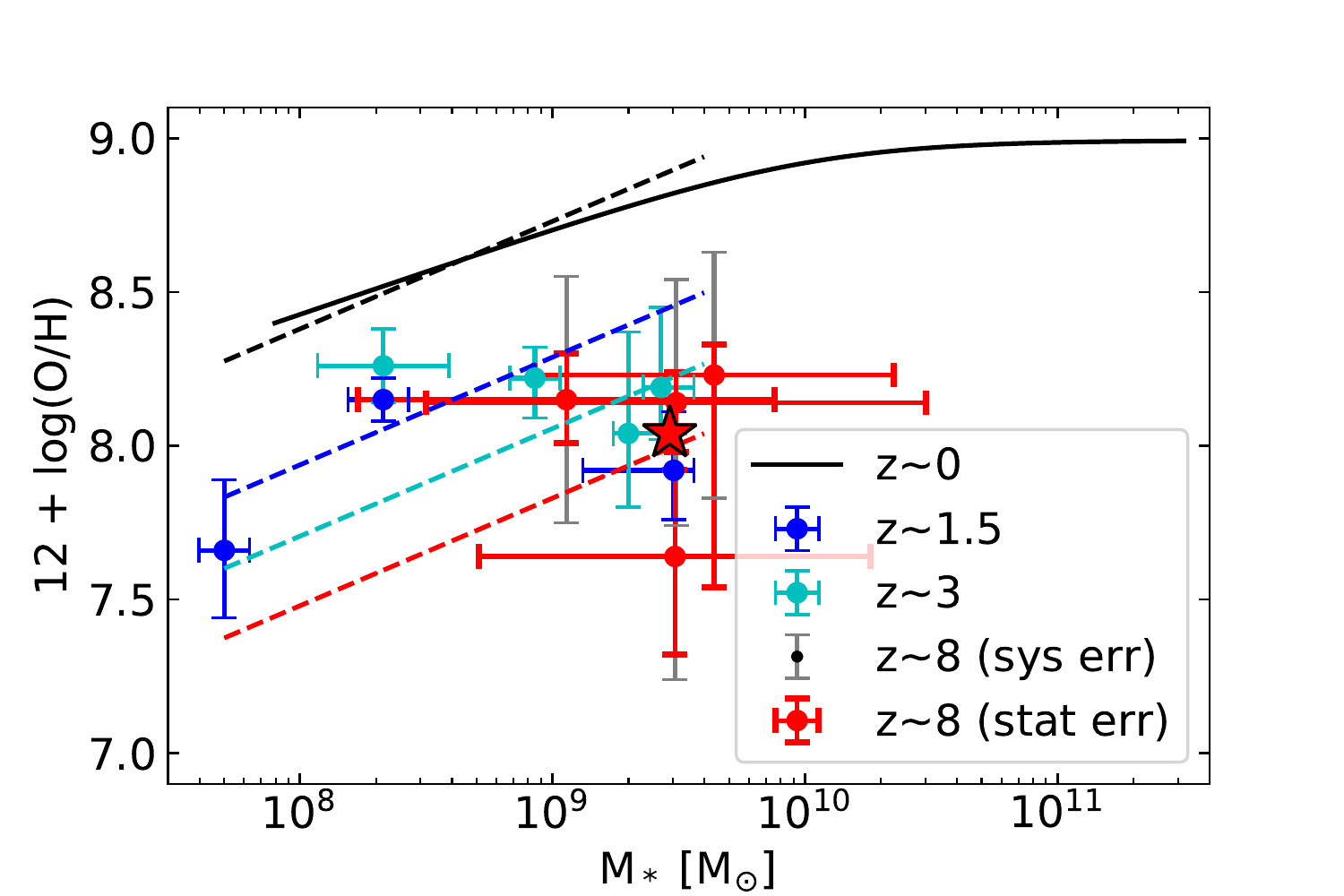}
}
\caption{
\label{fig:MZR}
Mass-metallicity relation of $z>7$ galaxies, compared with $z\simeq1.5$--3.5 galaxies with \Oiiit-based measurements listed in Table~\ref{tab:literature}, and the $z\simeq0$ relation from stacked spectra \citep[black line;][]{Curti2020}. Theoretical simulation-based results from \cite{Ma2016} are shown as dashed lines with the same redshift color-coding ($z=0$, 1.5, 3, and 8). \\
{\it Left:} Mass-metallicity relation with original stellar mass values reported in the literature. 
{\it Right:} Stellar masses for the $z>7$ sample from \citet[][summarized in Table~\ref{tab:literature}]{Roberts-Borsani2020}: error bars in mass span the 1- and 2-component SED fit results, while data points represent the midpoint of these two cases. Metallicities are from this work (Table~\ref{tab:metallicity}) and are the same in both panels. 
The galaxy SXDF-NB1006-2 is not shown in the right panel as it was not included in the sample of \cite{Roberts-Borsani2020}, due to insufficient mid-IR photometry. Metallicities from the literature are increased by 0.2 dex in the right panel, showing the approximate effect of temperature fluctuations which can systematically underestimate \Oiiit-based values. 
The abundance scale is such that the most massive $z\simeq0$ galaxies have moderately super-solar metallicities (by $\sim$0.1 and 0.3 dex in the left and right panels, respectively, where we adopt solar \OH~$=8.69$ from \citealt{Asplund2009}). 
In both panels, red error bars represent statistical measurement uncertainty for the $z>7$ sample, grey error bars show the estimated 0.4 dex systematic uncertainty in O/H, and red stars are the $z>7$ unweighted mean. 
}
\end{figure*}

We show the mass-metallicity relation (MZR) in Figure~\ref{fig:MZR} for two different sets of stellar masses reported in the literature (summarized in Table~\ref{tab:literature}). Masses in the left panel are based on fits which suggest ages $\lesssim10$ Myr for most sources. It is perhaps better to treat these as lower limits following the discussion above \citep[e.g.,][]{Inoue2016}. The right panel shows the mean and range of masses for two limiting cases from \cite{Roberts-Borsani2020}, where allowing for extended star formation histories can increase the stellar mass by a factor of $\sim30\times$. 
(We note that SXDF-NB1006-2 was not included in the sample of \citealt{Roberts-Borsani2020} due to insufficient mid-IR photometry, and is not shown in the right panel of Figure~\ref{fig:MZR}.)
Here we consider the implications for metallicity. As discussed above, the young and low-mass case generally requires very large \Oiii$\lambda\lambda$4959,5007 fluxes to reproduce observed mid-IR colors. Such strong line emission implies high \Te\ and $\mathrm{O_{32}}$ (hence lower ICF) compared to our fiducial values (Sections~\ref{sec:sample} and \ref{sec:uncertainties}). As a result our metallicity estimates would be revised downward by $\sim$0.2 dex. The high-mass extreme corresponds to an opposite effect with much weaker W$_{\mathrm{[O III] 4959,5007 + H\beta}} \lesssim 200$~\AA, such that metallicity estimates would be revised upward. 
Since the low/high mass cases would imply lower/higher metallicity, even $\pm0.8$ dex in mass would shift the sample by $\lesssim0.2$ dex in O/H relative to the theoretical mass-metallicity relation of \cite{Ma2016}. 
In this sense the uncertainty in MZR evolution is plausibly smaller than uncertainty in metallicity alone. 
Otherwise this mass range would propagate to $\pm0.3$ dex in MZR evolution for the conservative case of treating mass and metallicity independently. 
Our interpretation of MZR evolution from the right panel of Figure~\ref{fig:MZR} is therefore not strongly affected by the large uncertainties in stellar mass.

Although the current sample is insufficient to measure the MZR slope at $z\simeq8$, it provides a useful basis to examine redshift evolution and compare with theoretical work. In this sense the relevant quantity is the (unweighted) mean \OH~$=7.9$, at stellar mass $\Mstar\sim 10^{9}-10^{9.5}~\Msun$. The comparison with \Te-based metallicities at lower redshift is also subject to systematic uncertainty arising from the \Oiiit\ diagnostic, which exhibits an ADF of $\sim$0.2 dex compared to other methods \citep[as noted in Section~\ref{sec:sample}, e.g.,][]{Esteban2009,Blanc2015}. Hence in the right panel of Figure~\ref{fig:MZR} we increase all lower-redshift values by 0.2 dex in O/H. 
This increase brings the local MZR to \OH~$\approx9.0$ for massive galaxies ($\Mstar \gtrsim 10^{10.5}~\Msun$), closer to most results in the literature \citep[e.g.,][]{Tremonti2004,Kewley2008,Mannucci2010}. 
We note that joint analyses of optical and far-IR lines in $z\simeq0$ galaxies are needed to fully address this discrepancy. At fixed $\Mstar$, the resulting evolution in O/H from $z=8\rightarrow0$ is $\sim$0.6 and 0.9 dex for the two panels in Figure~\ref{fig:MZR}. 
We consider the latter to be a more appropriate relative comparison, and thus estimate the evolution as $0.9\pm0.5$ dex in O/H accounting conservatively for systematic uncertainty in both metallicity and $\Mstar$. 

Several groups have reported theoretical predictions for the mass-metallicity relation at high redshifts based on cosmological simulations. Bright $z\simeq8$ galaxies near the mass range of our sample are predicted to have metallicities near $\sim$0.1 solar \citep{Moriwaki2018,Katz2019,Langan2020}, with evolution in the MZR to $z=8$ of $\sim$0.5--0.9 dex \citep[Figure~\ref{fig:MZR};][]{Ma2016,Torrey2019}. 
These predictions are compatible and in fact in good agreement with our results. A potentially interesting point of divergence is the MZR evolution from $z\simeq4-8$, where predictions range from $\sim$0--0.3 dex \citep{Langan2020}, reflecting differences in galaxy gas content and metal-enriched outflows. 
Precise metallicities from future JWST and ALMA data (Section~\ref{sec:jwst}) can usefully test these models. Chemical evolution studies can therefore address the formative processes of gas accretion and stellar feedback in reionization-era galaxies.

\subsection{Spatially resolved metallicity}

The sensitivity and angular resolution of ALMA (and JWST) permits resolved metallicity measurements on $\sim$1 kpc scales at $z>7$, using the approach outlined herein. Resolved structure is already evident with ALMA in many cases \citep[e.g.,][]{Hashimoto2019,Smit2018,Matthee2017}. Such information can provide considerable insight into the earliest phases of galaxy assembly, for example regarding in-situ growth versus hierarchical merging. Distinguishing mergers from disk galaxies can be challenging even with high-quality kinematic data \citep[e.g.,][]{Simons2019}, while resolved metallicity can be a powerful discriminant \citep[e.g.,][]{Wang2017,Wang2020}. For in-situ galaxy growth, spatial variations in metallicity provide insight into gas mixing and feedback processes \citep[e.g.,][]{Ma2017,Jones2013,Wang2019}. 

BDF-3299 is a compelling example of resolved structure in our sample. Our fiducial case gives limits \OH~$<7.9$ and $>8.1$ for the UV- and IR-bright components respectively. Likewise B14-65666 is resolved into two distinct clumps, where luminosities reported by \cite{Hashimoto2019} suggest that clump ``A'' is more metal-rich by a few tenths of a dex than clump ``B'' for our fiducial case. These nominal differences could indicate mergers in both systems, where the more enriched component likely dominates the stellar mass. Although the metallicities of each component are not distinguishable given the uncertainties, it illustrates the prospects of resolving chemical enrichment and metal mixing at $z>7$. This is a truly remarkable possibility in our view, especially as no such resolved \Te\ measurements have yet been achieved beyond the relatively nearby universe.

\section{Removing systematic uncertainties: $<$0.2 dex precision with JWST and ALMA}\label{sec:jwst}

Having discussed implications of our measurements for chemical enrichment in the reionization epoch, we now consider prospects for improved precision with future JWST and ALMA data. While prospects for stellar mass and strong-line abundance methods have been widely recognized \citep[e.g.,][]{Roberts-Borsani2020}, here we focus on \Te-based metallicity using the far-IR emission lines and \Oiiif\ in particular. 

The results of Section~\ref{sec:uncertainties} provide both an assessment of the dominant uncertainties and guidance on how to minimize them. Each source of systematic uncertainty can be reduced or eliminated with a modest number of emission line measurements, summarized in Table~\ref{tab:uncertainty}. We frame the discussion below in approximate order of the largest to smallest sources of uncertainty, leaving ICF for last as it depends on \Te\ and \Ne.

\begin{enumerate}

\item $L$(\Hb): JWST can provide both \Hb\ flux and the correction for dust attenuation (from ratios such as \Ha/\Hb\ or \Pa/\Hb). Attenuation correction is also critical for \Te\ (via \Oiii$\lambda\lambda$4959,5007) and ICF (via \Oiid). 
We note that Paschen series lines are accessible to JWST/MIRI at $z>7$ and are more robust than Balmer lines to the dust attenuation curve. The \Pa\ feature is especially promising, having approximately 10$\times$ lower attenuation than \Hb. 

\item \Ne: The most promising approach in our view is to measure \Oiiif, accessible in the 450 $\mu$m atmospheric window with ALMA (Band 9) for $z\simeq7.1-8.6$. With $n_{crit}=3470$ \Nunit, \Oiiif\ is nearly independent of density for normal \Hii\ region conditions, while the \Oiiif/\Oiiie\ ratio provides the density \Ne(\Opp). Figure~\ref{fig:density} illustrates the clear advantage of \Oiiif\ for robust O/H measurements. Density diagnostics such as the optical \Sii\ and \Oii\ doublets are also accessible to JWST, although they do not directly probe the \Opp\ ionization state. 

\item \Te: \Oiii$\lambda$5007 and/or $\lambda$4959 luminosity provides \Te\ in combination with the far-IR lines, ideally \Oiiif. We view the \Oiiit\ temperature diagnostic as far less pragmatic than \Oiiif,88$\mu$m given its low expected flux (see Section~\ref{sec:4363}). 

\item ICF (\Op\ and \Oppp\ abundance): \Oiid\ is the most accessible feature to address the ICF, although we consider some lingering uncertainty from unknown \Te(\Op). 
A range $\pm1000$ K ($\pm3000$ K) in \Te(\Op) corresponds to only $\pm0.03$ dex ($\pm0.1$ dex) in $\log{\mathrm{(O/H)}}$ for the fiducial conditions. The magnitude of ICF uncertainty is related to \Oii\ flux, such that higher $\mathrm{O_{32}}$ translates to smaller uncertainty. \Oppp\ is expected to contribute minimally ($\sim$1\%; Section~\ref{sec:ICF}), although it should be reconsidered in cases where \Op/\Opp\ is very low ($\lesssim0.05$). 

\end{enumerate}

The four points above outline a clear path toward removing systematic uncertainties using a modest set of strong emission lines: \Oiiif, \Oiii$\lambda$5007, \Hb, other \Hi\ lines such as \Pa\ and/or \Ha, and \Oiid\ (Table~\ref{tab:uncertainty}). 
To avoid aperture correction uncertainties, rest-optical fluxes are best determined with grism (WFSS: wide field slitless spectroscopy) or integral field spectroscopy (IFS) modes of JWST. 
With these features, systematic uncertainty is limited by the ICF and estimated to be $\lesssim$0.05-0.1 dex in O/H.

\subsection{Feasibility of JWST and ALMA observations}

In this section we comment on expected emission line fluxes and detection feasibility. \Hi\ lines will likely present the greatest challenge for JWST. For the brighter sources B14-65666 and SXDF-NB1006-2, with a conservative $A_V=1$ magnitude of dust extinction, the luminosity estimates from Table~\ref{tab:metallicity} translate to expected fluxes $f_{H\beta} \simeq 9\times10^{-18}~\funit$, $f_{H\alpha} \simeq 3 \times10^{-17}$, and $f_{Pa\alpha} \simeq 7\times10^{-18}~\funit$. \Hb\ and \Ha\ would be $\sim$2 times stronger for estimates of $A_V\lesssim0.3$ \citep{Hashimoto2019,Inoue2016}, while \Pa\ is relatively insensitive to $A_V$. Thus the brighter objects are expected to have \Hi\ line fluxes $\gtrsim10^{-17}~\funit$. \Oiii$\lambda$5007 is likely to be stronger by a factor of a few, based on both empirical and theoretical estimates \citep[e.g.,][]{Roberts-Borsani2020,Katz2019}. \Oiid\ may be weaker than \Hb, but uncertainty in \Oii\ flux does not necessarily limit the precision in O/H \citep[Section~\ref{sec:forecast};][]{Belli2013}. 
JWST IFS and WFSS sensitivity estimates suggest that all of these key features can be detected at $\gtrsim10\sigma$ significance within a few hours or less, for the brighter sources. We note that grism spectroscopy offers comparable or superior spatial resolution and sensitivity over a far larger field of view than IFS, which may even be useful for multiplexing of clustered $z>7$ sources. JWST's grism modes thus provide a powerful method to probe the relevant nebular emission lines (e.g. \Oii, \Hb, and \Oiii$\lambda$5007 at $z>7$ with NIRCam WFSS). 

Finally, the key \Oiiif\ line is comparable or brighter than \Oiiie\ (depending on \Ne; Figure~\ref{fig:density}). The minimum flux ratio is $\frac{f_{52}}{f_{88}} = 0.6$ for low densities, increasing to 1.8 for \Ne~$=600$~\Nunit\ (and 1.1 for the fiducial case \Ne~$=250$~\Nunit). 
The main challenge is that \Oiiif\ falls in ALMA's high-frequency bands, requiring excellent conditions and of order $\sim$10-hour integrations for the brightest objects. From the sample in Table~\ref{tab:metallicity}, those at $z<8$ have \Oiiif\ at relatively accessible frequencies. 
We note that compact ALMA array configurations are optimal for this work, but are typically available only during poor-weather months, with little time suitable for high-frequency observations. Scheduling compact configurations during months with reliable high-frequency weather (e.g. May through August) would greatly improve the feasibility of observing \Oiiif\ at high redshifts with ALMA.

\subsection{Forecasted precision in oxygen abundance}\label{sec:forecast}

To illustrate the potential of combining rest-optical and far-IR emission lines, we perform a calculation of \Te, \Ne, and \OH\ with full error propagation. We adopt 10\% flux uncertainty for \Hb, \Pa, and \Oiiie; 20\% flux uncertainty for \Oiiif\ and \Oiid; and $\pm1000$ K uncertainty in \Te(\Op) from the $T_2-T_3$ relation (Section~\ref{sec:ICF}). Line fluxes are chosen to give the fiducial \Te~$=1.5\times10^4$~K, \Ne~$=250$~\Nunit, ICF~=~0.17 dex, with 0.15 solar abundance and $A_V=1$ magnitude (although varying $A_V$ does not affect the results). The total result is \OH~$=7.88\pm0.17$, with \Oiiif\ dominating the error budget ($\pm0.15$ dex). The other substantial error source is \Pa\ ($\pm0.07$; the same precision is obtained for \Ha\ with 3\% flux uncertainty). \Hb, \Oiid, and \Te(\Op) each contribute $\pm0.03$ while \Oiiie\ contributes negligibly ($<0.01$ dex, illustrating mild effects of \Ne). \Oiii$\lambda$5007 likewise contributes negligible error, assuming a flux sensitivity comparable to \Hb. Increasing the flux uncertainty to 20\% for all lines results in $\pm0.23$ dex uncertainty in O/H.

We conclude that signal-to-noise ratios (SNR) $\gtrsim5$ with good flux calibration are sufficient to achieve $\lesssim0.2$ dex precision in O/H. 
The limiting factors above are \Oiiif\ and \Pa, which anchor the \Opp\ and \Hp\ abundances. The latter can be reasonably improved with higher sensitivity and multiple \Hi\ lines (e.g. simultaneously measured with JWST/MIRI-MRS). \Oiiif\ is likely to be limited by flux calibration accuracy of $\sim$10\%, corresponding to a best-case precision of $\pm0.08$ dex achievable with deep ALMA observations.

\subsection{Comparison of \Oiiif\ and \Oiiit}\label{sec:4363}

In Section~\ref{sec:52um} we argue that the far-IR offers considerable advantages over optical auroral lines at high redshifts. For comparison, we repeat the analysis of Section~\ref{sec:forecast} using only rest-optical emission lines, with \Te\ based on \Oiiit. For the same physical properties, a flux uncertainty of 25\% in \Oiiit\ (i.e., 4$\sigma$ detection) propagates to \OH~$=7.88^{+0.17}_{-0.12}$ assuming infinite SNR in all other lines. This precision is comparable to the case outlined in Section~\ref{sec:forecast}. 
However, \Oiiit\ is 12 times weaker than \Hb\ and 60 times weaker than \Oiii$\lambda$5007 for our fiducial conditions (e.g., \Te~$=1.5\times10^4$~K). Hence achieving SNR=4 in \Oiiit\ requires $\gtrsim$10 times better sensitivity at rest-optical wavelengths ($\sim$100$\times$ longer integration times with JWST!) compared to achieving SNR=5 in \Hb, and may be even more challenging if \Te\ is lower. We therefore find the \Oiiif\ approach with ALMA to be far more pragmatic.

\section{Discussion}\label{sec:summary} 

We present a simple method to determine gas-phase oxygen abundance from far-IR emission lines and SFR measurements, and apply it to a sample of galaxies at $z=7.1-9.1$.\footnote{Following the submission of this work, \cite{Yang2020} presented an analytic model for \Oiii\ fine structure emission based on similar physical arguments. They verify their model with CLOUDY calculations, and reproduce our metallicity estimates from Equation 2 to within 0.03 dex, deriving consistent metallicities for the sample discussed herein. This provides further validation of our methodology with an independent framework.} 
 We find typical \OH~$=7.9$ or $\sim$0.2 times the solar abundance. We determine evolution in the MZR of $0.9\pm0.5$ dex in O/H from $z=8\rightarrow0$ at fixed stellar mass, accounting conservatively for systematic uncertainties in both mass and metallicity. 
The results reveal a substantial degree of chemical enrichment in reionization-era galaxies, indicating that star formation has likely been ongoing for $\sim$100 Myr in much of the sample \citep[supporting multi-component studies of stellar population properties; e.g.,][]{Roberts-Borsani2020}. The underlying evolved stellar populations are not captured by single-component photometric fits, yet can easily dominate the stellar mass and integrated number of UV photons produced by these galaxies. This in turn is important for understanding the cosmic reionization process at $z>7$.

Notably, \cite{Roberts-Borsani2020} constrain the extent of older stellar populations largely using \Oiiie\ and far-IR dust continuum. These ALMA measurements enable modeling of how nebular emission and stellar continuum breaks contribute to observed mid-IR photometry. 
Additional higher-frequency ALMA data can help to constrain both the dust mass and nebular emission (as \Oiiif\ would eliminate the factor of 3$\times$ uncertainty from the range of densities considered herein, compared to \Oiiie). 
While JWST spectroscopy presents a clear path forward, further ALMA observations are a promising intermediate step to refine estimates of stellar population properties.

We have discussed prospects for improving metallicity measurements with future JWST and ALMA data, concluding that 0.1--0.2 dex precision is feasible with modest integration times for the best $z>7$ targets. The \Oiiif\ emission line is highlighted as an important feature for securing robust \Te\ and abundance measurements. While challenging high-frequency observations are required, we nonetheless argue that \Oiiif\ is far more accessible than the \Oiiit\ \Te\ diagnostic. 
While our discussion focuses on $z>7$ galaxies, the same methods can be applied at lower $z=5.2$--6.3 where \Oiiif\ is redshifted into ALMA's Band 10. ALMA's sensitivity to fixed \Oiiif\ luminosity is approximately equal in Band 9 ($z\approx7$--8) as in Band 10, as the $\sim2\times$ higher noise is compensated by lower luminosity distance. Intrinsically luminous (or gravitationally lensed) $z\simeq6$ galaxies are thus detectable within a few hours with ALMA in high-frequency weather. 
As noted in Section~\ref{sec:jwst}, these measurements would ideally make use of future ALMA cycles with compact arrays scheduled during seasons with reliable high-frequency weather.

Future abundance measurements can provide remarkable insight into the early assembly history of galaxies. 
Knowledge of metallicity will improve photoionization models used to interpret high-redshift emission lines, including inferences on the stellar ionizing spectrum. Metallicity also serves as a constraint on integrated enrichment from past star formation history, especially in combination with improved stellar mass determinations. Star formation histories and the stellar ionizing spectrum in turn have important implications for the timeline of cosmic reionization and the role of star forming galaxies. 
Additionally, measuring evolution in the mass-metallicity relation from $z\simeq5$--8 will test different theoretical models and provide insight into the gas content, accretion rates, and stellar feedback which drive early galaxy evolution. Current theoretical models vary by $\sim$0.3 dex of evolution in O/H, which can be distinguished with modest samples using the techniques described herein. 
Finally, spatially resolved metallicity measurements may be feasible for bright and moderately extended sources, offering a useful probe of gas mixing processes and hierarchical merger assembly. 
JWST is undoubtedly poised to revolutionize our knowledge of reionization-era galaxies, and we view ALMA observations of far-IR \Oiii\ lines as an extremely valuable component for chemical evolution studies.

\begin{deluxetable*}{lcccccccc}
\tablecaption{Physical properties of the Herschel DGS sample\label{tab:DGS}}
\tablehead{
Object name & Distance\tablenotemark{a} & 12+log(O/H)\tablenotemark{a} & 12+log(O/H)\tablenotemark{b} & $\Mstar$\tablenotemark{a} & SFR\tablenotemark{a} & \Oiiie\ flux\tablenotemark{c} & $L$(\Oiiie)\tablenotemark{d} & 12+log(O/H)\tablenotemark{e} 
 \\
  &  [Mpc]  &  (PT05)  &  (I06)  &  [$10^8 \, \Msun$]  &  [$\Msunyr$]  &  [$10^{-18}$~W\,m$^{-2}$]  &  [$10^{7} \, \Lsun$]  &  (Eq.~\ref{eq:Opp_abundance})  
}
\medskip
\startdata

Haro 11 & 92.1 & 8.36 & 8.23 & 339.0 & 28.6 & 1720.0 & 45.62 & 8.11 \\
Haro 2 & 21.7 & 8.23 & --- & 20.8 & 0.9 & 972.0 & 1.432 & 8.11 \\
Haro 3 & 19.3 & 8.28 & 8.37 & 18.8 & 0.8 & 1850.0 & 2.156 & 8.34 \\
He 2-10 & 8.7 & 8.43 & --- & 18.2 & 0.79 & 3380.0 & 0.8003 & 7.91 \\
HS 1222+3741 & 181.7 & 7.79 & 7.83 & 14.6 & 1.507 & 14.6 & 1.508 & 7.91 \\
HS 1304+3529 & 78.7 & 7.93 & 7.66 & 4.2 & 0.5 & 31.9 & 0.618 & 8.00 \\
HS 1330+3651 & 79.7 & 7.98 & 7.66 & 6.3 & 0.382 & 29.8 & 0.5923 & 8.09 \\
II Zw 40 & 12.1 & 8.23 & 8.09 & 4.7 & 0.43 & 3590.0 & 1.645 & 8.48 \\
Mrk 1450 & 19.8 & 7.84 & 8.03 & 0.86 & 0.47 & 262.0 & 0.3213 & 7.74 \\
Mrk 930 & 77.8 & 8.03 & 8.09 & 44.2 & 3.12 & 421.0 & 7.973 & 8.31 \\
NGC 1140 & 20.0 & 8.38 & 8.27 & 23.4 & 0.57 & 1080.0 & 1.352 & 8.28 \\
NGC 5253 & 4.0 & 8.25 & 8.16 & 6.7 & 0.24 & 9010.0 & 0.451 & 8.18 \\
Pox 186 & 18.3 & 7.70 & 7.75 & 0.1 & 0.04 & 33.7 & 0.0353 & 7.85 \\
SBS 1159+545 & 57.0 & 7.44 & 7.46 & 0.5 & 0.17 & 4.81 & 0.0489 & 7.36 \\
UGC 4483 & 3.2 & 7.46 & 7.55 & 0.03 & 0.0007 & 25.7 & 0.000823 & 7.98 \\
UM 133 & 22.7 & 7.82 & 7.70 & 0.81 & 0.02 & 24.5 & 0.0395 & 8.20 \\
UM 448 & 87.8 & 8.32 & 8.03 & 241.0 & 12.36 & 1060.0 & 25.56 & 8.22 \\
UM 461 & 13.2 & 7.73 & 7.79 & 0.26 & 0.01 & 79.0 & 0.04306 & 8.54 \\
VII Zw 403 & 4.5 & 7.66 & 7.74 & 0.1 & 0.003 & 81.6 & 0.00517 & 8.14 

\enddata
\tablenotetext{a}{As reported in Table~2 of \citet{Madden2013}; references for distance and metallicity are given therein. Metallicity is calculated using the method of \citet[][abbreviated as PT05]{Pilyugin2005}.}
\tablenotetext{b}{As reported in Table~6 of \citet{Madden2013}. Metallicity is calculated using the method of \citet[][abbreviated as I06]{Izotov2006}. Missing values indicate cases where \Oiiit\ was not available, and therefore a direct-method metallicity measurement was not possible. }
\tablenotetext{c}{As reported in Table~4 of \citet{Cormier2015}.}
\tablenotetext{d}{Calculated using the listed values of distance and \Oiiie\ flux.}
\tablenotetext{e}{Calculated using the listed values of \Oiiie\ luminosity and SFR via Equation~\ref{eq:Opp_abundance}, with $+0.17$ dex ionization correction. }
\end{deluxetable*}

\begin{deluxetable*}{lccccccc}
\tablewidth{0pt}
\tablecaption{Oxygen abundance estimates of $z>7$ galaxies\label{tab:metallicity}}
\tablehead{
\colhead{Object name\tablenotemark{a}} & \colhead{$z_{spec}$\tablenotemark{a}} & \colhead{$L$(\Oiiie)\tablenotemark{a}} & \colhead{SFR\tablenotemark{a}} & \colhead{$L$(\Hb)\tablenotemark{b}} & \colhead{\OHpp} & \colhead{\OH\tablenotemark{c}} & References\tablenotemark{d}
 \\
                      &                      & \colhead{[$\Lsun$]}   & \colhead{[$\Msunyr$]} & \colhead{[erg~s$^{-1}$]} & \colhead{$\pm \sigma_{stat}$} & \colhead{$\pm \sigma_{stat} \pm \sigma_{sys}$} & 
}
\medskip
\startdata

MACS1149-JD1\tablenotemark{e} & 9.110 & $(7.4\pm1.6)\times10^{7}$ & 4.2$^{+0.8}_{-1.1}$ & $3.43\times10^{41}$ & 7.98$^{+0.15}_{-0.14}$ & 8.15$^{+0.15}_{-0.14} \pm 0.4$ & H18, L19 \\
A2744-YD4\tablenotemark{e} & 8.382 & $(7.0\pm1.7)\times10^{7}$ & 12.9$^{+11.1}_{-6.0}$ & $1.05\times10^{42}$ & 7.47$^{+0.28}_{-0.32}$ & 7.64$^{+0.28}_{-0.32} \pm 0.4$ & L17, L19 \\
MACS0416-Y1\tablenotemark{e} & 8.312 & $(1.2\pm0.3)\times10^{9}$ & 57$^{+175.0}_{-0.2}$ & $4.65\times10^{42}$ & 8.06$^{+0.10}_{-0.69}$ & 8.23$^{+0.10}_{-0.69} \pm 0.4$ & T19 \\
SXDF-NB1006-2 & 7.215 & $(9.9\pm2.1)\times10^{8}$ & 219$^{+105}_{-176}$ & $1.79\times10^{43}$ & 7.39$^{+0.71}_{-0.21}$ & 7.56$^{+0.71}_{-0.21} \pm 0.4$ & I16 \\
B14-65666 & 7.168 & $(3.4\pm0.4)\times10^{9}$ & 200$^{+82}_{-38}$ & $1.63\times10^{43}$ & 7.97$^{+0.10}_{-0.16}$ & 8.14$^{+0.10}_{-0.16} \pm 0.4$ & H19, F16 \\
BDF-3299 (UV)\tablenotemark{f} & 7.109 & $<5.5\times10^{7}$ & $>$5.7 & $>4.65\times10^{41}$ & 7.72 & 7.89 & C17, M15 \\
BDF-3299 (IR)\tablenotemark{f} & 7.109 &  $(1.8\pm0.2)\times10^{8}$ & $<$12  & $<9.80\times10^{41}$ & 7.91 & 8.08 & C17, M15 

\enddata
\tablenotetext{a}{As compiled in Table~2 of \citet{Harikane2020}.}
\tablenotetext{b}{\Hb\ luminosity calculated from SFR via Equation~\ref{eq:L_Ha}.}
\tablenotetext{c}{\OH\ includes a fiducial correction of $+0.17$ dex from singly-ionized oxygen.}
\tablenotetext{d}{References for redshift, \Oiiie\ luminosity, and SFR. C17: \citet{Carniani2017}, F16: \citet{Furusawa2016}, H18: \citet{Hashimoto2018}, H19: \citet{Hashimoto2019}, I16: \citet{Inoue2016}, L17: \citet{Laporte2017}, L19: \citet{Laporte2019}, M15: \citet{Maiolino2015}, T19: \citet{Tamura2019}
}
\tablenotetext{e}{Luminosity and SFR values are corrected for lensing magnification. All other quantities are independent of magnification.}
\tablenotetext{f}{Far-IR line emission is spatially offset from the rest-UV continuum,  
and we give results separately for these two spatial locations. 
For the rest-UV component we list the reported upper limit on $L$(\Oiiie) (which assumes a line width of 100~$\kms$), and the reported SFR assuming no extinction (which we express as a lower limit). 
For the IR \Oiiie-emitting component we list the reported 2-$\sigma$ upper limit on SFR \citep[although this limit may be underestimated, as noted by][]{Carniani2017}. 
We refer readers to \cite{Carniani2017} for further details of the measurements, limits, and spatial structure of this system. 
Oxygen abundances correspond to the listed values in both cases and should be treated as an upper (lower) limit for the UV (IR) region. Given the complexities of upper and lower limits, we do not attempt to quantify the uncertainty range.}
\end{deluxetable*}

\begin{deluxetable*}{lccccl}
\tablewidth{0pt}
\tablecaption{Sources of systematic uncertainty in oxygen abundance\label{tab:uncertainty}}
\tablehead{
\colhead{Parameter} & \colhead{Fiducial value} & \colhead{Range} & \colhead{$\sigma_{\log{\mathrm{(O/H)}}}$\tablenotemark{a}} & \colhead{$\frac{\partial \log{\mathrm{(O/H)}}}{\partial \log{X}}$\tablenotemark{b}} & \colhead{Prospects for eliminating systematic error at $z>7$} 
}
\medskip
\startdata

\Te  &  15,000 K  &  9,000--20,000 K  &  $^{+0.17}_{-0.09}$  & $-0.72$ &  \Oiii$\lambda\lambda$4959,5007 with JWST, \& \Oiiif\ with ALMA  \\

\Ne  &  250 \Nunit  &  10--600 \Nunit  &  $^{+0.23}_{-0.24}$  & $0.47$ &  \Oiiif\ with ALMA  \\

$L$(\Hb)  &  (Equation~\ref{eq:L_Ha})  &  RMS~$\simeq$~0.25 dex  &  $\pm0.25$  & $-1.0$ &  \Hb\ and other \Hi\ lines with JWST  \\

$\mathrm{\frac{O^+ ~+~ O^{3+}}{O_{tot}}}$  &  0.33  &  0.08--0.59  &  $^{+0.22}_{-0.13}$  & $0.25$ &  \Oii$\lambda\lambda$3727,3729 with JWST  

\enddata
\tablenotetext{a}{Here $\sigma$ corresponds to the difference in $\log{\mathrm{(O/H)}}$ between the fiducial value and extrema of the range.}
\tablenotetext{b}{Partial derivative of $\log{\mathrm{(O/H)}}$ with respect to each parameter ($X$: \Te, \Ne, etc.), evaluated at the fiducial values.}
\end{deluxetable*}

\begin{deluxetable*}{lcccccc}
\tablewidth{0pt}
\tablecaption{Literature mass and metallicity measurements\label{tab:literature}}
\tablehead{
\colhead{Object name} & \colhead{$z$} & \colhead{\OH}  & \colhead{$\log{\frac{\Mstar}{\Msun}}$\tablenotemark{a}} & \colhead{$\log{\frac{\Mstar}{\Msun}\tablenotemark{b}}$} & \colhead{$\log{\frac{\Mstar}{\Msun}\tablenotemark{b}}$}  &  References  \\
                      &               & (\Oiiit-based) &   & (R-B20, 1-comp) & (R-B20, 2-comp) 
}
\medskip
\startdata

 & & &  $z>7$ sample  &  \\
\hline

MACS1149-JD1          &      9.110    &                & $9.03^{+0.17}_{-0.08}$ & $8.23\pm0.01$ & $9.88\pm0.01$   &  H18   \\
A2744-YD4             &      8.382    &                & $9.29^{+0.24}_{-0.18}$ & $8.71\pm0.02$ & $10.26\pm0.06$  &  L17  \\
MACS0416-Y1           &      8.312    &                & $8.38^{+0.10}_{-0.06}$ & $8.93\pm0.01$ & $10.35\pm0.03$  &  T19   \\
SXDF-NB1006-2         &      7.215    &                & $8.54^{+0.79}_{-0.22}$ &    &        &  I16  \\
B14-65666             &      7.168    &                & $8.89^{+0.05}_{-0.04}$ & $9.25\pm0.04$ & $9.49\pm0.99$   &  H19   \\

\hline
 & & & $z=1-4$ sample &  \\
\hline

S13\tablenotemark{c}          &      1.425    &  $7.95^{+0.07}_{-0.07}$  &  $8.33^{+0.1}_{-0.14}$  \\
AEGIS-11452\tablenotemark{c}  &      1.6715   &  $7.72^{+0.19}_{-0.16}$  &  $9.48^{+0.08}_{-0.36}$  \\
C12a\tablenotemark{c}         &      1.8339   &  $7.46^{+0.23}_{-0.22}$  &  $7.7^{+0.1}_{-0.1}$  \\
GOODS-S-41547\tablenotemark{c}&      2.5451   &  $7.84^{+0.33}_{-0.24}$  &  $9.3^{+0.13}_{-0.06}$  \\
A1689-217\tablenotemark{d}    &      2.5918   &  $8.06\pm0.12$  &  8.07--8.59  \\
COSMOS-1908\tablenotemark{c}  &      3.0767   &  $8.02^{+0.10}_{-0.13}$  &  $8.93^{+0.1}_{-0.1}$  \\
COSMOS-23895\tablenotemark{c} &      3.6372   &  $7.99^{+0.26}_{-0.17}$  &  $9.43^{+0.13}_{-0.07}$  

\enddata
\tablenotetext{a}{References for stellar mass are given in the final column, with same codes as Table~\ref{tab:metallicity}.}
\tablenotetext{b}{1- and 2-component masses from R-B20: \cite{Roberts-Borsani2020}. Masses are corrected for lensing magnification, where relevant.}
\tablenotetext{c}{As compiled by \cite{Sanders2020}.}
\tablenotetext{d}{From \cite{Gburek2019}.}
\end{deluxetable*}

\section*{ACKNOWLEDGEMENTS}

We thank Brian Lemaux and Xin Wang for helpful comments and discussions. 
We thank the referee for providing a constructive report which improved the content and clarity of this manuscript. 
We are grateful to the authors of PyNeb, and related packages FIVEL and nebular, for providing reliable code which forms the basis of our analysis. 
RSE acknowledges funding from the European Research Council (ERC) under the European Union’s Horizon 2020 research and innovation program (grant agreement No. 669253). 
TT acknowledges support by NSF through grant AST-1810822; TT and GRB acknowledge support by NASA through grant JWST-ERS-01324.001-A. 
TJ and RS acknowledge support by NASA through grant \#07-0182 issued by the Universities Space Research Association, Inc. 
The results herein make use of previously reported data from several ALMA programs. ALMA is a partnership of ESO (representing its member states), NSF (USA) and NINS (Japan), together with NRC (Canada), MOST and ASIAA (Taiwan), and KASI (Republic of Korea), in cooperation with the Republic of Chile. The Joint ALMA Observatory is operated by ESO, AUI/NRAO and NAOJ. 



\begin{thebibliography}{43}
\expandafter\ifx\csname natexlab\endcsname\relax\def\natexlab#1{#1}\fi

\bibitem[Andrews \& Martini(2013)]{Andrews2013} Andrews, B.~H., \& Martini, P.\ 2013, \apj, 765, 140

\bibitem[Asplund et al.(2009)]{Asplund2009} Asplund, M., Grevesse, N., Sauval, A.~J., et al.\ 2009, \araa, 47, 481

\bibitem[Belli et al.(2013)]{Belli2013} Belli, S., Jones, T., Ellis, R.~S., \& Richard, J.\ 2013, \apj, 772, 141

\bibitem[Blanc et al.(2015)]{Blanc2015} Blanc, G.~A., Kewley, L., Vogt, F.~P.~A., et al.\ 2015, \apj, 798, 99

\bibitem[Bouwens et al.(2019)]{Bouwens2019} Bouwens, R.~J., Stefanon, M., Oesch, P.~A., et al.\ 2019, \apj, 880, 25

\bibitem[Calvi et al.(2016)]{Calvi2016} Calvi, V., Trenti, M., Stiavelli, M., et al.\ 2016, \apj, 817, 120

\bibitem[Campbell et al.(1986)]{Campbell1986} Campbell, A., Terlevich, R., \& Melnick, J.\ 1986, \mnras, 223, 811

\bibitem[Carniani et al.(2017)]{Carniani2017} Carniani, S., Maiolino, R., Pallottini, A., et al.\ 2017, \aap, 605, A42

\bibitem[Christensen et al.(2012)]{Christensen2012} Christensen, L., Laursen, P., Richard, J., et al.\ 2012, \mnras, 427, 1973 

\bibitem[Cormier et al.(2015)]{Cormier2015} Cormier, D., Madden, S.~C., Lebouteiller, V., et al.\ 2015, \aap, 578, A53

\bibitem[Croxall et al.(2013)]{Croxall2013} Croxall, K.~V., Smith, J.~D., Brandl, B.~R., et al.\ 2013, \apj, 777, 96

\bibitem[Cullen et al.(2019)]{Cullen2019} Cullen, F., McLure, R.~J., Dunlop, J.~S., et al.\ 2019, \mnras, 487, 2038

\bibitem[Curti et al.(2020)]{Curti2020} Curti, M., Mannucci, F., Cresci, G., et al.\ 2020, \mnras, 491, 944

\bibitem[Ellis et al.(2013)]{Ellis2013} Ellis, R.~S., McLure, R.~J., Dunlop, J.~S., et al.\ 2013, \apjl, 763, L7

\bibitem[Esteban et al.(2009)]{Esteban2009} Esteban, C., Bresolin, F., Peimbert, M., et al.\ 2009, \apj, 700, 654


\bibitem[Faisst et al.(2016)]{Faisst2016} Faisst, A.~L., Capak, P.~L., Davidzon, I., et al.\ 2016, \apj, 822, 29

\bibitem[Furusawa et al.(2016)]{Furusawa2016} Furusawa, H., Kashikawa, N., Kobayashi, M.~A.~R., et al.\ 2016, \apj, 822, 46

\bibitem[Gburek et al.(2019)]{Gburek2019} Gburek, T., Siana, B., Alavi, A., et al.\ 2019, \apj, 887, 168

\bibitem[Guseva et al.(2012)]{Guseva2012} Guseva, N.~G., Izotov, Y.~I., Fricke, K.~J., et al.\ 2012, \aap, 541, A115

\bibitem[Harikane et al.(2020)]{Harikane2020} Harikane, Y., Ouchi, M., Inoue, A.~K., et al.\ 2020, \apj, 896, 93

\bibitem[Hashimoto et al.(2018)]{Hashimoto2018} Hashimoto, T., Laporte, N., Mawatari, K., et al.\ 2018, \nat, 557, 392

\bibitem[Hashimoto et al.(2019)]{Hashimoto2019} Hashimoto, T., Inoue, A.~K., Mawatari, K., et al.\ 2019, \pasj, 71, 71

\bibitem[Huang et al.(2016)]{Huang2016} Huang, K.-H., Brada{\v{c}}, M., Lemaux, B.~C., et al.\ 2016, \apj, 817, 11

\bibitem[Inoue et al.(2016)]{Inoue2016} Inoue, A.~K., Tamura, Y., Matsuo, H., et al.\ 2016, Science, 352, 1559

\bibitem[Izotov et al.(1994)]{Izotov1994} Izotov, Y.~I., Thuan, T.~X., \& Lipovetsky, V.~A.\ 1994, \apj, 435, 647

\bibitem[Izotov et al.(2006)]{Izotov2006} Izotov, Y.~I., Stasi{\'n}ska, G., Meynet, G., Guseva, N.~G., \& Thuan, T.~X.\ 2006, \aap, 448, 955 

\bibitem[Izotov et al.(2014)]{Izotov2014} Izotov, Y.~I., Thuan, T.~X., \& Guseva, N.~G.\ 2014, \mnras, 445, 778

\bibitem[Je{\v{r}}{\'a}bkov{\'a} et al.(2018)]{Jerabkova2018} Je{\v{r}}{\'a}bkov{\'a}, T., Hasani Zonoozi, A., Kroupa, P., et al.\ 2018, \aap, 620, A39

\bibitem[Jones et al.(2013)]{Jones2013} Jones, T., Ellis, R.~S., Richard, J., \& Jullo, E.\ 2013, \apj, 765, 48

\bibitem[Jones et al.(2015)]{Jones2015} Jones, T., Martin, C., \& Cooper, M.~C.\ 2015, \apj, 813, 126

\bibitem[Katz et al.(2019)]{Katz2019} Katz, H., Galligan, T.~P., Kimm, T., et al.\ 2019, \mnras, 487, 5902

\bibitem[Kennicutt(1998)]{Kennicutt1998} Kennicutt, R.~C.\ 1998, \araa, 36, 189

\bibitem[Kewley \& Ellison(2008)]{Kewley2008} Kewley, L.~J., \& Ellison, S.~L.\ 2008, \apj, 681, 1183

\bibitem[Komatsu et al.(2011)]{Komatsu2011} Komatsu, E., Smith, K.~M., Dunkley, J., et al.\ 2011, \apjs, 192, 18

\bibitem[Labb{\'e} et al.(2013)]{Labbe2013} Labb{\'e}, I., Oesch, P.~A., Bouwens, R.~J., et al.\ 2013, \apjl, 777, L19

\bibitem[Lagos et al.(2018)]{Lagos2018} Lagos, P., Scott, T.~C., Nigoche-Netro, A., et al.\ 2018, \mnras, 477, 392

\bibitem[Langan et al.(2020)]{Langan2020} Langan, I., Ceverino, D., \& Finlator, K.\ 2020, \mnras, 494, 1988

\bibitem[Laporte et al.(2014)]{Laporte2014} Laporte, N., Streblyanska, A., Clement, B., et al.\ 2014, \aap, 562, L8

\bibitem[Laporte et al.(2017)]{Laporte2017} Laporte, N., Ellis, R.~S., Boone, F., et al.\ 2017, \apjl, 837, L21

\bibitem[Laporte et al.(2019)]{Laporte2019} Laporte, N., Katz, H., Ellis, R.~S., et al.\ 2019, \mnras, 487, L81

\bibitem[L{\'o}pez-S{\'a}nchez et al.(2012)]{Lopez-Sanchez2012} L{\'o}pez-S{\'a}nchez, {\'A}. R., Dopita, M.~A., Kewley, L.~J., et al.\ 2012, \mnras, 426, 2630

\bibitem[Luridiana et al.(2012)]{Luridiana2012} Luridiana, V., Morisset, C., \& Shaw, R.~A.\ 2012, IAU Symposium, 422

\bibitem[Ly et al.(2014)]{Ly2014} Ly, C., Malkan, M.~A., 
Nagao, T., et al.\ 2014, \apj, 780, 122 

\bibitem[Ma et al.(2016)]{Ma2016} Ma, X., Hopkins, P.~F., Faucher-Gigu{\`e}re, C.-A., et al.\ 2016, \mnras, 456, 2140

\bibitem[Ma et al.(2017)]{Ma2017} Ma, X., Hopkins, P.~F., Feldmann, R., et al.\ 2017, \mnras, 466, 4780

\bibitem[Madden et al.(2013)]{Madden2013} Madden, S.~C., R{\'e}my-Ruyer, A., Galametz, M., et al.\ 2013, \pasp, 125, 600

\bibitem[Mainali et al.(2020)]{Mainali2020} Mainali, R., Stark, D.~P., Tang, M., et al.\ 2020, \mnras, 494, 719

\bibitem[Maiolino et al.(2015)]{Maiolino2015} Maiolino, R., Carniani, S., Fontana, A., et al.\ 2015, \mnras, 452, 54

\bibitem[Maiolino \& Mannucci(2019)]{Maiolino2019} Maiolino, R., \& Mannucci, F.\ 2019, \aapr, 27, 3

\bibitem[Mannucci et al.(2010)]{Mannucci2010} Mannucci, F., Cresci, G., Maiolino, R., et al.\ 2010, \mnras, 408, 2115

\bibitem[Mason et al.(2018)]{Mason2018} Mason, C.~A., Treu, T., Dijkstra, M., et al.\ 2018, \apj, 856, 2

\bibitem[Matthee et al.(2017)]{Matthee2017} Matthee, J., Sobral, D., Boone, F., et al.\ 2017, \apj, 851, 145

\bibitem[Moriwaki et al.(2018)]{Moriwaki2018} Moriwaki, K., Yoshida, N., Shimizu, I., et al.\ 2018, \mnras, 481, L84

\bibitem[Oesch et al.(2018)]{Oesch2018} Oesch, P.~A., Bouwens, R.~J., Illingworth, G.~D., et al.\ 2018, \apj, 855, 105

\bibitem[Palay et al.(2012)]{Palay2012} Palay, E., Nahar, S.~N., Pradhan, A.~K., et al.\ 2012, \mnras, 423, L35

\bibitem[Pilyugin \& Thuan(2005)]{Pilyugin2005} Pilyugin, L.~S. \& Thuan, T.~X.\ 2005, \apj, 631, 231

\bibitem[Roberts-Borsani et al.(2016)]{Roberts-Borsani2016} Roberts-Borsani, G.~W., Bouwens, R.~J., Oesch, P.~A., et al.\ 2016, \apj, 823, 143

\bibitem[Roberts-Borsani et al.(2020)]{Roberts-Borsani2020} Roberts-Borsani, G.~W., Ellis, R.~S., \& Laporte, N.\ 2020, \mnras, 497, 3440

\bibitem[Sanders et al.(2016)]{Sanders2016} Sanders, R.~L., Shapley, A.~E., Kriek, M., et al.\ 2016, \apj, 816, 23

\bibitem[Sanders et al.(2020)]{Sanders2020} Sanders, R.~L., Shapley, A.~E., Reddy, N.~A., et al.\ 2020, \mnras, 491, 1427

\bibitem[Sargsyan \& Weedman(2009)]{Sargsyan2009} Sargsyan, L.~A. \& Weedman, D.~W.\ 2009, \apj, 701, 1398

\bibitem[Scoville et al.(2017)]{Scoville2017} Scoville, N., Lee, N., Vanden Bout, P., et al.\ 2017, \apj, 837, 150

\bibitem[Shapley et al.(2017)]{Shapley2017} Shapley, A.~E., Sanders, R.~L., Reddy, N.~A., et al.\ 2017, \apjl, 846, L30

\bibitem[Shivaei et al.(2016)]{Shivaei2016} Shivaei, I., Kriek, M., Reddy, N.~A., et al.\ 2016, \apjl, 820, L23

\bibitem[Simons et al.(2019)]{Simons2019} Simons, R.~C., Kassin, S.~A., Snyder, G.~F., et al.\ 2019, \apj, 874, 59

\bibitem[Smit et al.(2014)]{Smit2014} Smit, R., Bouwens, R.~J., Labb{\'e}, I., et al.\ 2014, \apj, 784, 58

\bibitem[Smit et al.(2015)]{Smit2015} Smit, R., Bouwens, R.~J., Franx, M., et al.\ 2015, \apj, 801, 122

\bibitem[Smit et al.(2018)]{Smit2018} Smit, R., Bouwens, R.~J., Carniani, S., et al.\ 2018, \nat, 553, 178

\bibitem[Stark(2016)]{Stark2016} Stark, D.~P.\ 2016, \araa, 54, 761

\bibitem[Stasi{\'n}ska(1982)]{Stasinska1982} Stasi{\'n}ska, G.\ 1982, \aaps, 48, 299

\bibitem[Strait et al.(2020)]{Strait2020} Strait, V., Brada{\v{c}}, M., Coe, D., et al.\ 2020, \apj, 888, 124

\bibitem[Strom et al.(2017)]{Strom2017} Strom, A.~L., Steidel, C.~C., Rudie, G.~C., et al.\ 2017, \apj, 836, 164

\bibitem[Tamura et al.(2019)]{Tamura2019} Tamura, Y., Mawatari, K., Hashimoto, T., et al.\ 2019, \apj, 874, 27

\bibitem[Torrey et al.(2019)]{Torrey2019} Torrey, P., Vogelsberger, M., Marinacci, F., et al.\ 2019, \mnras, 484, 5587

\bibitem[Totani et al.(2006)]{Totani2006} Totani, T., Kawai, N., Kosugi, G., et al.\ 2006, \pasj, 58, 485

\bibitem[Tremonti et al.(2004)]{Tremonti2004} Tremonti, C.~A., Heckman, T.~M., Kauffmann, G., et al.\ 2004, \apj, 613, 898

\bibitem[Vincenzo et al.(2016)]{Vincenzo2016} Vincenzo, F., Matteucci, F., Belfiore, F., et al.\ 2016, \mnras, 455, 4183

\bibitem[Wang et al.(2017)]{Wang2017} Wang, X., Jones, T.~A., Treu, T., et al.\ 2017, \apj, 837, 89

\bibitem[Wang et al.(2019)]{Wang2019} Wang, X., Jones, T.~A., Treu, T., et al.\ 2019, \apj, 882, 94

\bibitem[Wang et al.(2020)]{Wang2020} Wang, X., Jones, T.~A., Treu, T., et al.\ 2020, arXiv e-prints, arXiv:1911.09841

\bibitem[Woosley \& Weaver(1995)]{Woosley1995} Woosley, S.~E., \& Weaver, T.~A.\ 1995, \apjs, 101, 181

\bibitem[Yang \& Lidz(2020)]{Yang2020} Yang, S. \& Lidz, A.\ 2020, arXiv:2007.14439

\bibitem[Zhang et al.(2018)]{Zhang2018} Zhang, Z.-Y., Ivison, R.~J., George, R.~D., et al.\ 2018, \mnras, 481, 59


\end{thebibliography}
\end{document}